\documentclass[aps,prapplied,reprint,superscriptaddress]{revtex4-2}

\usepackage{graphicx}% Include figure files
\usepackage{dcolumn}% Align table columns on decimal point
\usepackage{bm}% bold math
\usepackage{xcolor}
\usepackage{gensymb}
\usepackage{subcaption}
\usepackage{amsmath}
\usepackage{comment}
\usepackage{booktabs}

\begin{document}
\preprint{APS/123-QED}
%\title{Robust Optical Tweezers Configuration for Sub-micron force Measurements}% Force line breaks with \\
%%1. Off-axis retro-reflected optical trap for surface force sensing with levitated particles
%%2. 3D Optical Trapping and Cooling near a

%AD:"A new method of optical trapping at sub-micron distances to a reflective surface for ultra-sensitive force measurements."
%AD: "Optical Trapping and feedback cooling of a nanoparticle at distinct subwavelength lattice sites near a conducting surface" 

%AG: "Optical trapping and feedback cooling of a nanoparticle in an angled standing wave trap near a conducting surface"
%\title{Angled optical standing wave trap for sub-micron force measurements with nanospheres near a conducting surface}
%\title{An angled optical standing wave trap for nanospheres at micron-range from a mirror}
\title{A method for optically trapping nanospheres at micron range from a tilted mirror}
%AAG: I am aprehensive about putting force sensing in the title if we don't report any force sensing in the manuscript, there is enough novelty here in the new trap geometry I think, and we can point out the potential for sensing.

%\author{Alexey Grinin\footnotemark{*}, Andrew Dana\footnotemark{*}, Mark Nguyen, Eduardo Alejandro, Andrew A. Geraci}
%\affiliation{Center for Fundamental Physics, Northwestern University.}%Lines break automatically or can be forced with \\
\author{Alexey Grinin}
\altaffiliation{These authors contributed equally.}
\affiliation{Center for Fundamental Physics, Department of Physics and Astronomy, Northwestern University, Evanston, Illinois 60208, USA}

\author{Andrew Dana}
\altaffiliation{These authors contributed equally.}
\affiliation{Center for Fundamental Physics, Department of Physics and Astronomy, Northwestern University, Evanston, Illinois 60208, USA}

\author{Mark Nguyen}
\affiliation{Center for Fundamental Physics, Department of Physics and Astronomy, Northwestern University, Evanston, Illinois 60208, USA}

\author{Eduardo Alejandro}
\affiliation{Center for Fundamental Physics, Department of Physics and Astronomy, Northwestern University, Evanston, Illinois 60208, USA}

\author{Andrew A. Geraci}
\affiliation{Center for Fundamental Physics, Department of Physics and Astronomy, Northwestern University, Evanston, Illinois 60208, USA}
\date{\today}% It is always \today, today,
\begin{abstract}
    %We propose and experimentally demonstrate a novel optical method for reliably loading, trapping, and laser cooling dielectric nanospheres at (sub)-micron distances from a reflective metallic surface. \textcolor{red}{By introducing an off-axis reflective surface near a single-beam optical trap, the interference between the incident and reflected beams creates a three-dimensional optical potential along their symmetry axis.} Stable potential minima emerge within a finite overlap region close to the surface, with their number, shape, and distance from the surface tunable via the incidence angle and polarization of the incoming beam. This configuration enables deterministic selection of trapping sites as the system transitions from a single-beam trap to a three-dimensional trap. We validate this approach by launching a 170nm-diameter silica sphere into a single-beam trap and transitioning it into the second potential minimum, located 1.64 µm from the surface. Additionally, we perform parametric feedback cooling of all three motional degrees of freedom in a high-vacuum environment. The experimental results align well with our theoretical model, supported by numerical simulations of the Langevin equations of motion. This method provides a robust platform for ultra-sensitive scanning surface force sensing and may open new pathways for short-range gravity measurements.

    We propose and experimentally demonstrate a novel optical method for trapping and cooling dielectric nanospheres at (sub)-micron distances from a reflective metallic surface. By translating a tilted mirror towards the focus of a single-beam optical tweezer, the optical trap transitions into an off-axis standing-wave configuration due to interference between the incident and reflected beams. Stable potential minima emerge within a finite overlap region close to the surface, with their number, shape, and distance from the surface tunable via the incidence angle, waist, and polarization of the incoming beam. This configuration enables deterministic selection of trapping sites as the system transitions from the single-beam trap to the off-axis standing wave trap. We validate this approach using a $170$ nm diameter silica sphere in a single-beam trap with a $1.5$ $\mu$m waist and transitioning it into the second or first potential minimum of the standing wave trap, located $1.61$ $\mu$m or $0.55$ $\mu$m from the surface, respectively. The experimental results align well with our theoretical model, supported by numerical simulations of the Langevin equations of motion. Additionally, we perform \textcolor{black}{optical parametric feedback cooling and electrostatic feedback cooling of all three motional degrees of freedom in a high-vacuum environment}. This method provides a robust platform for ultra-sensitive scanning surface force sensing \textcolor{black}{with neutral and charged particles }at micron distances from a reflective surface in high vacuum and may open new pathways for short-range gravity or Casimir effect measurements.
\end{abstract}
\maketitle

%\tableofcontents

\section{\label{sec:intro}Introduction}
%#########################################################################################################################################
%#########################################################################################################################################

%\textit{Introduction: Andy}: 

Over the past several years, levitated optomechanical systems have emerged as a promising platform for precision measurements of weak forces and studies of macroscopic quantum mechanics \cite{Millen_2020,reviewmarkus}. This is due largely to the extreme environmental decoupling made possible by suspending nano- or micro-particles in a high vacuum environment using optical radiation pressure or electromagnetic fields. For example, mechanical quality factors in excess of $10^{10}$ have been demonstrated with dielectric nanoparticles suspended in a Paul trap at ultra-high vacuum \cite{Tracy2024}. Several groups have recently demonstrated ground state cooling of the center of mass oscillations of optically levitated nanospheres in optical cavities \cite{delic2020cooling,Marin2022} or in free space using electrostatic feedback cooling \cite{tebbenjohanns2021quantum,Novotny2022, groundstate, groundstatecryo}. 
Optically levitated particles in high vacuum have been demonstrated as sensors of extremely small forces, e.g., of order $10^{-21}$N \cite{Ranjit:2016}, as well as feeble accelerations \cite{andyhart2015, Moore2017,novotnydrop}, torques \cite{Li2016}, and rotations \cite{Li2018,Novotny2018,Moore2018}. Levitated optomechanics has also been identified as a testbed for foundational aspects of quantum mechanics \cite{oriol2011}, observing quantum behavior in mechanical systems \cite{chang2009,coherentscattering, aspelmeyercavity}, and as a tool for quantum information science \cite{sympcool}. 

For force sensing, one possibility is to study the interactions of a suspended nanoparticle with a nearby surface \cite{Geraci:2010, Montoya:22,Magrini:18,Novotnymembrane, Winstone2018}. Such systems are promising for fundamental physics investigations such as searching for non-Newtonian deviations in the gravitational force at micron-scale distances as suggested by string theory or other physics beyond the standard model, or for studying the Casimir effect \cite{Geraci:2010}. Another potential practical application is scanning force microscopy for electric field measurements or magnetic field measurements for appropriately functionalized, e.g., charged or magnetic, nanoparticles.
In order to conduct measurements of weak forces with such systems, it is essential to have a robust method of repeatably placing a levitated sensor a known distance away from a source while also suppressing any background noise which makes measurements of weak forces inaccessible. This is particularly relevant for short range investigations of gravitational forces, where background electromagnetic forces need to be screened using conducting surfaces.

In this article, we describe a method for introducing a nanoparticle near a reflective conducting mirror tilted approximately at a $45$ degree angle.  
%To date, mitigating such background noises has proven difficult, although the use of conductive surfaces as shields has become a widely used method for such applications. Previous experiments using conductive surfaces as shields have been successful in bringing microspheres distances around tens of microns to such a surface\textcolor{red}{check these numbers}. However, the capability to repeatably place a levitated sensor sub-micron distances from a conducting surface is still a sought after system.  \textcolor{red}{maybe don't include backaction stuff since we didn't really investigate this so much...}In the context of macroscopic quantum mechanics, ground state cooling of the motional degrees of freedom of levitated nanospheres in vacuum has been achieved.  Looking towards the future of such systems, it is crucial to study the characteristics of backaction noise resulting from random perturbations in the laser field which limits the achievable measurement imprecision. The ability to mitigate such measurement backaction and reduce measurement imprecision noise will expand the possible applications of levitated nanoparticles as macroscopic quantum systems.  
%In this article, we present 
The apparatus is capable of placing a levitated sensor at a small number of well defined micron and sub-micron distances away from the surface with the ability to tune the center-of-mass frequencies %and dipole radiation pattern 
\textit{in-situ}. By using a highly focused Gaussian beam to trap a 170 nm diameter silica nanosphere, it is possible to transition this particle into an optical lattice formed by the reflection of the trapping beam from a tilted conducting surface by slowly moving this surface towards the trap with a nanopositioning stage. We demonstrate reasonable agreement between our experimental results and our theoretical model of the optical trapping configuration, including particle simulations using the Langevin equation of motion.
% and potential methods of studying backaction suppression for macroscopic quantum mechanical experiments. 

%To place this method in context with previous approaches that have been investigated, in Ref. \cite{Montoya:22} a method was demonstrated to load a dielectric nanoparticle into the anti-node of a retro-reflected standing wave near a conducting surface. However in this case a thin reflecting mirror was inserted into an optical tweezer trap just behind the location where a silica nanosphere is confined, so that the particle can be loaded into an anti-node of the retro-reflected standing wave trap formed near the surface. 
In  previously demonstrated optical standing wave traps near a conducting surface, such as in Ref. \cite{Montoya:22}, where a thin reflecting mirror is inserted into an optical tweezer beam just behind the location of a trapped nanoparticle, there is more difficulty in placing a nanoparticle in the same lattice site each time one is trapped, as the position of the mirror surface relative to the trapped particle needs to be adjusted with sub-micron level precision.  By using a highly focused beam incident on a tilted reflective surface, the number of lattice sites where a particle can be stably trapped can be greatly reduced.  The system described here provides only two stable lattice sites which have greatly different potential depths.  Due to the vast reduction in possible lattice sites and the distinct characteristics of each, it is possible to accurately determine the distance of the levitated sensor from the conducting surface by simply measuring the center of mass frequencies, an already commonplace and often necessary experimental method in levitated optomechanical experiments. This eliminates the need to implement additional cumbersome mechanisms such as additional lasers or fiber interferometers for determining the distance of the particle to the surface. Furthermore, the tilted mirror approach we describe allows observation of the displacement of the particle from both the forwards and backwards scattered light, as the forwards directed scattered light can be collected in a similar fashion to that used in single beam optical tweezer traps. 

Finally, as needed for future ultra-sensitive force detection applications, we demonstrate three-dimensional parametric laser feedback cooling of the particle at micron-range from the surface, needed to stabilize the nanoparticle for force sensing in high vacuum.  \textcolor{black}{To highlight the compatibility with other state of the art cooling methods, we additionally demonstrate electrostatic three dimensional linear feedback cooling of the particles mechanical degrees of freedom.} These results represent a possible new method to facilitate precision measurements of weak forces such as gravity and Casimir-Polder forces in the sub-micron regime using \textcolor{black}{neutral and charged} optically trapped nanospheres.  %environment and the ability to experimentally measure the distance of the particle to the surface in one of the lattice sites. \textcolor{red}{need to add something else here and make the intro a bit more tidy and concise...have not yet mentioned the usefulness of being able to have a lattice trap which still allows for interferometric detection using the forward scattered light and the backwards scattered light.}

%#########################################################################################################################################
%#########################################################################################################################################

\section{\label{sec:theory}Theoretical Background}

The optical trapping potential in our system is generated by the interference between a tightly focused incident Gaussian beam and its reflection from a tilted mirror surface. To model this system, we employ a modified Gaussian beam approximation \textcolor{black}{(see Appendix B for details)}, where the transverse beam waists $w_u$, $w_v$, and the Rayleigh range $z_R$ are treated as independent parameters. Here, the $u$-axis corresponds to the beam’s polarization direction, while the $v$-axis corresponds to the orthogonal polarization direction. These parameters are fitted to match the more rigorous Debye diffraction integral solution \cite{Novotny-Hecht-2006}, which fully accounts for the vectorial nature and nonparaxial effects of tightly focused fields. Although the Debye integral provides a more accurate description, the modified Gaussian beam model offers substantial analytical insight and remains a good approximation even for relatively high numerical apertures. In particular, for our objective with NA = 0.67, this approach yields good agreement with the Debye diffraction solutions while greatly simplifying the analysis of the optical potential as further detailed in Appendix B.

For a Gaussian TEM$_{00}$ mode propagating along the $z$-direction reflected off a mirror at 45 degrees (see Fig. \ref{fig:setup}a and \ref{fig:potentialcontour1}) with waists $w_{0u}$, $w_{0v}$, Rayleigh length $z_R$, and electric field amplitude $E_0$, the optical potential is given by:
\onecolumngrid
\begin{equation}
U(\vec{r}) =  -\frac{\alpha'}{4} \left\{
\vec{\mathcal{E}}_{\text{inc}}^2(\vec{r}) + \vec{\mathcal{E}}_{\text{ref}}^2( \vec{r})
- 2 \vec{\mathcal{E}}_{\text{inc}}(\vec{r})\cdot\vec{\mathcal{E}}_{\text{ref}} (\vec{r})
\cos\!\Bigl[ k(z + y) + \tfrac{k(x^2 + y^2)}{2 R(z)} - \tfrac{k(x^2 + z^2)}{2 R(y)} - \xi(z) + \xi(y) \Bigr]
\right\}.
\end{equation}
\twocolumngrid

Here $\alpha'$ is the real part of the polarizability of a dielectric nanosphere $\alpha = \alpha'+i\alpha''=\alpha_0\bigl(1-i\alpha_0 k^3/6\pi\epsilon_0\bigr)^{-1}$ and $\alpha_0=3V\epsilon_0(n^2-1)/(n^2+2)$ (Clausius–Mossotti relation). The field envelope amplitudes are given by:
\begin{eqnarray}
\vec{\mathcal{E}}_{\text{inc}} &=& E_0\frac{w_{0u}\,w_{0v}}{w_u(z)\,w_v(z)}\,
\exp\Bigl[- \tfrac{x^2}{w_u^2(z)} - \tfrac{y^2}{w_v^2(z)}\Bigr] \,\hat{x}, 
\\[6pt]
\vec{\mathcal{E}}_{\text{ref}} &=& E_0\frac{w_{0u}\,w_{0v}}{w_u(y)\,w_v(y)}\,
\exp\Bigl[- \tfrac{x^2}{w_u^2(y)} - \tfrac{z^2}{w_v^2(y)}\Bigr] \,\hat{x}.
\end{eqnarray}
We refer to the waist along the polarization direction of a beam as its $u$-axis and the orthogonal polarization direction as $v$-axis. Here, the beam waists along each axis evolve with distance as
\[
w_u(s) = w_{0u}\,\sqrt{1 + \bigl(s/z_R\bigr)^2},
\quad
w_v(s) = w_{0v}\,\sqrt{1 + \bigl(s/z_R\bigr)^2},
\]
while the wavefront curvature is given by $R(s) = s\bigl(1 + (z_R/s)^2\bigr)$ and the Gouy phase by $\xi(s) = \arctan\!\bigl(s/z_R\bigr)$. For an arbitrary angle between the incident and reflected beams, the expressions for the envelope amplitudes and phase terms can be generalized and are given in the appendix.
\begin{comment}
by projecting the spatial coordinate vector $\vec{r}$ along and orthogonal to each beam’s propagation direction. Specifically, the axial coordinates $z$ and $y$ in the expressions above should be replaced by $\vec{k}_{\text{inc}} \cdot \vec{r} / k$ and $\vec{k}_{\text{ref}} \cdot \vec{r} / k$, respectively, where $\vec{k}_{\text{inc}}$ and $\vec{k}_{\text{ref}}$ are the wavevectors of the incident and reflected beams. The transverse coordinates entering the Gaussian envelope then correspond to directions orthogonal to each beam axis. In this generalized form, the beam widths $w(s)$, radii of curvature $R(s)$, and Gouy phases $\xi(s)$ are defined along the effective propagation coordinate $s = \vec{k} \cdot \vec{r} / k$. If the beam waist is not located at the reflecting surface, the coordinates must be shifted by $\vec{r} \rightarrow \vec{r} - \vec{r}_0$, where $\vec{r}_0$ is the waist position, to accurately represent the field profile.
\newline
\end{comment}

When focus of the beam is near the surface (i.e.\ within the Rayleigh range), the amplitudes of the incident beam and its reflection are almost equal $\vec{\mathcal{E}}_{\text{inc}}\approx \vec{\mathcal{E}}_{\text{ref}}$ and the potential simplifies to:
\onecolumngrid
\begin{equation}
U(\vec{r}) \approx  -\frac{\alpha' E_0^2\,w_{0u}^2\,w_{0v}^2\,
\exp\!\Bigl[-2\Bigl(\tfrac{x^2}{w_u^2(z)}\;-\;\tfrac{y^2}{w_v^2(z)}\;-\;\tfrac{x^2}{w_u^2(y)}\;-\;\tfrac{z^2}{w_v^2(y)}\Bigr)\Bigr]}
{4\,w_u(z)\,w_v(z)\,w_u(y)\,w_v(y)}\;
\sin^2\!\Bigl[\tfrac{k}{2}\bigl(z + y + \tfrac{x^2 + y^2}{2R(z)} - \tfrac{x^2 + z^2}{2R(y)}\bigr) - \tfrac{1}{2}(\xi(z) - \xi(y))\Bigr].
\end{equation}
\twocolumngrid

The resulting 1D optical lattice is formed along the diagonal direction between the two beams, oriented at 45 degrees. The potential minima lie approximately along the line $y = -z$, where the phase terms satisfy $kz + ky = (2n+1)\pi$, if $k$ is the wave number and $n=0,1,2,\dots$ is an integer. This leads to potential wells located at $y = -z = (2n+1)\lambda/(4\sqrt{2})$, spaced by $\Delta z = \sqrt{2}\,\lambda/2$, where $\lambda$ is the laser wavelength. For tightly focused beams, only a small number of stable minima exist. For the beams used in this work, the Gouy phase and curvature terms shift the minima by just a few nanometers.

The time-averaged scattering force is obtained via
\begin{eqnarray}
\vec{F}_{\text{scatt}} &=& -\mu_0 \,\omega\,\alpha'' \,\langle \vec{S} \rangle 
= -\frac{\mu_0\,\alpha''}{2}\,E_0^2\,\nabla\phi
\nonumber \\[4pt]
&\approx& -\frac{\mu_0\,\alpha''}{2}\,E_0^2\,\frac{k}{\sqrt{2}}\,(\hat{z} - \hat{y}).
\end{eqnarray}
Here $\phi$ is the phase of the combined field, $\alpha''$ is the imaginary part of the particle’s polarizability, $\omega$ is the laser angular frequency, and $k$ is the wave number. The scattering force is directed along the sum vector of the beam propagation directions, i.e.\ parallel to the surface. It displaces the particle away from the potential minimum in that direction by tens of nanometers for the trap parameters and particles we consider. Notably, if two counter-propagating beams are symmetrically incident from opposing sides, the total scattering force cancels, enabling stable trapping even for larger particles.
%#########################################################################################################################################

\section{\label{sec:expmethods}Experimental Setup}
\begin{figure}[h!]
  \centering
  \includegraphics[width=\linewidth]{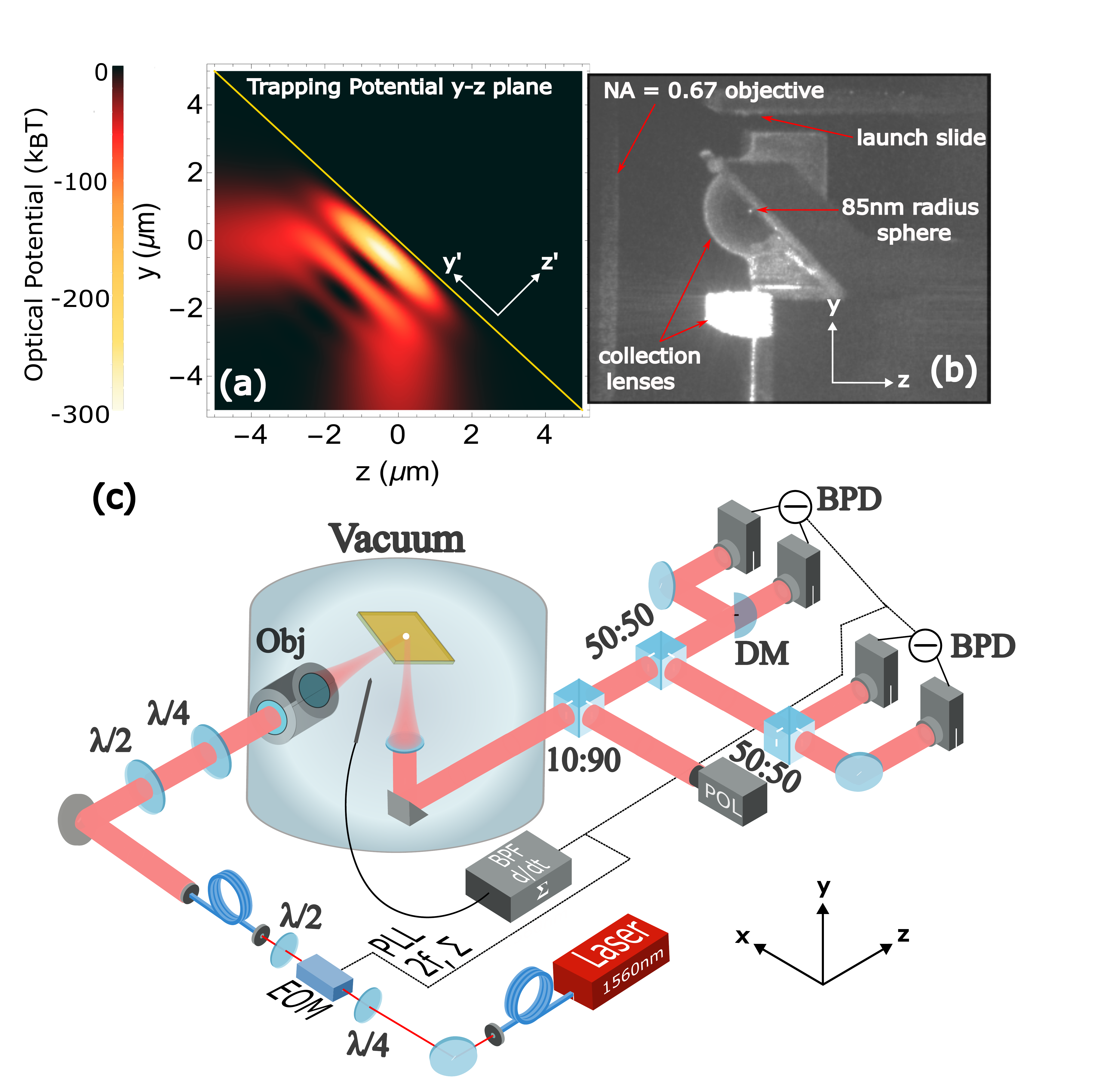}
  \caption{(a) Calculated optical potential in the $y-z$ plane for a TEM$_{00}$ Gaussian beam incident at $45$ degrees on a perfectly reflective surface.  The optical focus with waist $w_0 = 1.51$ $\mu$m is placed at the origin (at the reflective surface) with a laser power of $380$ mW and a temperature $T=300$K is used to have units of $k_B T$. (b) Image from a SWIR infrared camera of the $y-z$ plane displaying the trapping objective, collection lenses, piezo-launching substrate, angled reflective surface and a trapped sphere. (c) Diagram of the optical layout showing the two balanced interferometric photodetection setups in the forward scattered direction, \textcolor{black}{band pass and derivative filters for electrostatic cooling} and the phase locked loop parametric feedback cooling scheme using an electro-optic modulator.
}
  \label{fig:setup}
\end{figure}

The optical trap is created by using a microscope objective with numerical aperture (NA) of 0.67 (OptoSigma PAL-50-NIR-HR-LC00) and laser of wavelength $\lambda=1560$ nm (NKT Adjustik HP C15). This objective has an approximate $1/e^2$ waist $w_0 \approx 1.5$ $\mu$m as determined with a scanning slit beam profiler (Dataray Beam'R2). This larger working distance is preferential to provide the necessary space for collection optics and the tilted conducting surface. The optical set up is schematically represented in Fig. \ref{fig:setup}c. Spherical SiO$_2$ nanoparticles with a nominal diameter of $170$ nm and density \textcolor{black}{$\rho = 2.0$ g/cm$^3$} are initially trapped in a single beam optical tweezer at a pressure of $10$ Torr. \textcolor{black}{We note that the density of the nanoparticles used in this study is approximately 10 percent less than bulk silica due to the synthesis method utilized by the commercial vendor (Bangs Laboratories, Inc.), typical for nanospheres produced with the wet-based Stober method \cite{STOBER196862}.} 
The nanoparticles are launched into the trap by using a piezoelectric transducer to release nanoparticles from a glass substrate using the method described in Ref. \cite{weisman2022}\textcolor{black}{, some of which are trapped with a non-zero surface charge}.

\begin{figure*}[t]
    \begin{center}
    \includegraphics[width=\textwidth]{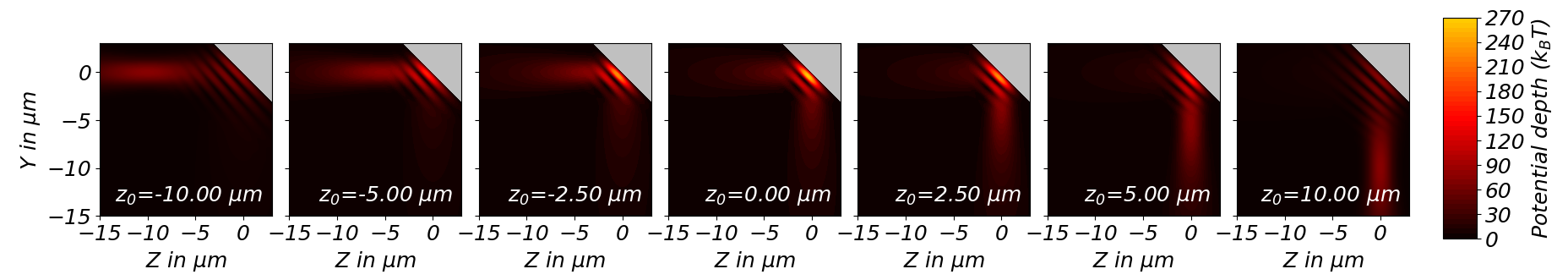}
    \caption{Contour plots of the optical potential depth generated by a TEM$_{00}$ Gaussian beam 
    ($w_{x0} = 1.36 \, \mu m$, $w_{y0} = 1.46 \, \mu m$, $P = 400$ mW) incident at 45 degrees on a mirror, 
    along with its reflected counterpart. The focal position is incrementally shifted, illustrating the 
    transition from an effectively single-beam configuration to a one-dimensional optical lattice. 
    The particle naturally follows this trajectory toward the most distant stable trapping site 
    (see Fig.~\ref{fig:adiabatic}), but it can be transferred to the next potential minimum 
    using resonant driving (compare Fig.~\ref{fig:jump}).}
    \label{fig:potentialcontour1}
    \end{center}
\end{figure*}
Once trapped, the pressure in the chamber is brought to $2$ Torr and the 45 degree reflective conducting surface, which is roughly $2$ mm away from the particle along the z-axis initially, is brought towards the particle via a nanopositioning stage (RS Scientific). As the optical focus approaches the reflective surface, a new optical potential is formed as shown in Fig. \ref{fig:setup}a and Fig. \ref{fig:potentialcontour1}.  This new optical potential is calculated using Eq. 1 and takes the focus to be at the surface. Additionally, Fig. \ref{fig:setup}b shows an image from a SWIR infrared camera depicting the trapping objective, launch slide, collection lenses, and tilted surface after it has begun moving towards the trapped particle.  The transition of the particle into the tilted lattice trap occurs when the single beam potential becomes weak enough compared to the nearby standing wave potential (See Fig. \ref{fig:potentialcontour1}).  Note that while the motional degrees of freedom in the single beam tweezer are defined along the lab frame axes denoted $\{x, y, z\}$ as labeled in Fig. \ref{fig:setup}c, once the particle has transitioned into the tilted lattice trap, the motion is along a new set of coordinates defined by the angle of the reflective surface relative to the incident beam. We denote this right handed coordinate system as $\{x', y', z'\}$ where the $z'$ axis is defined as normal to the reflective surface. The reflecting conductive surface (Norcada Inc.) is a $10$ mm$\times$$10$ mm$\times$$500$ $\mu$m Si frame with a $10$ nm Ti and $100$ nm Au coating.  Located at the center of the frame is a $2$ mm$\times$$2$ mm$\times$$150$ nm low stress SiN membrane also with the Au coating. \textcolor{black}{The Au coating is done via electron beam assisted evaporation. In this work only the thicker frame is used due to geometrical constraints associated with its mounting stage in the current setup.}

The polarization of the trapping light is chosen along the $x$-direction because this polarization state is unchanged upon reflection off of the tilted surface for the geometry shown here. In contrast, the orthogonal polarization state (y-direction) upon reflection becomes rotated by $90$ degrees resulting in zero interference.  This polarization dependence allows for \textit{in-situ} tunability of the potential depth (see Fig. \ref{fig:column_figures}) and distance to the surface. In practice, the polarization of the trapping beam is measured by picking off a small portion of light after the reflective surface and sending it to a polarimeter (Thorlabs PAX1000IR2).
%Furthermore, the transition of the particle from the single beam to the second standing wave potential is very robust and repeatable \textcolor{red}{somewhere earlier we must clearly define which wells are which}.  A variety of pressures and speeds have been tested, with the determining factor for successful transition being the polarization of the trapping beam which can cause insufficient potential depth to keep the particle trapped in the standing wave.  
%his occurs before the focus of the trapping beam is at the reflective surface, and thus there exists a regime where the COM frequencies in a given potential well can be changed in-situ by simply changing the relative distance between the particle and the focus.  This is done by translating the reflective surface mounted on a nanopositioning stage along the axis parallel to the optical axis of the trapping objective. 

\begin{figure*}[]
    \centering
    \includegraphics[width=\textwidth]{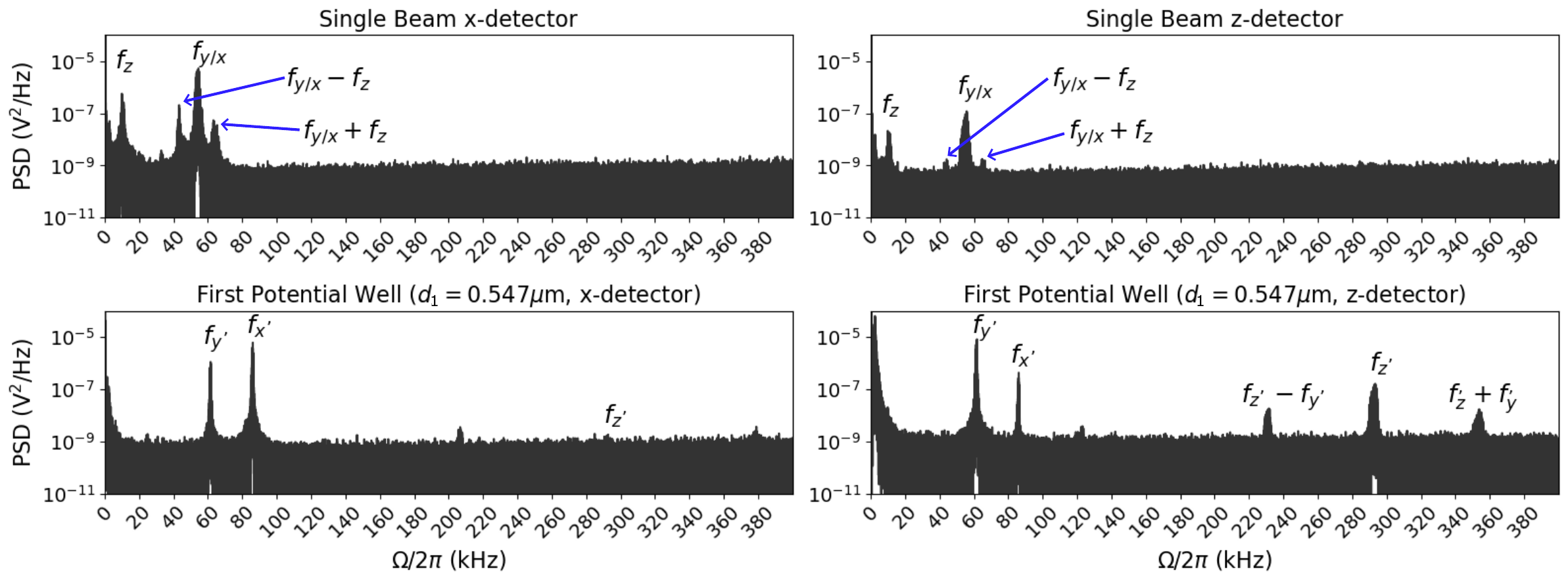}
    \caption{Power spectral density (PSD) from both detectors for a 170 nm diameter silica sphere trapped in the single beam tweezer (top row) and similarly for when the particle has been transitioned to the first potential well, which is calculated to be $0.547\mu$m from the reflective surface (bottom row). Note that in the first potential well, the frequencies for each degree of freedom are denoted with a prime as to distinguish them from the typical lab coordinates.}
   \label{fig:spectra1}
\end{figure*}

Measurement of center of mass motion for the single beam and the tilted lattice trap is done interferometrically using the forward scattered light from the particle and the trapping beam as the reference. From these measurements, a power spectral density can be determined as shown in Fig. \ref{fig:spectra1}. Detection of certain COM degrees of freedom can also be done using the backwards scattered light along with a separate local oscillator \cite{Tebben2019}.  In retro-reflected optical lattice traps for nanoparticles such as Ref. \cite{Montoya:22}, independent access to the forward and backward scattered light for interferometeric detection is not straight forward, whereas in the tilted configuration, information from the forward and backward scattered light are easily collected in similar fashion to typical optical tweezers. In this set up, the forward scattered light and trapping light are collected by a re-collimating lens below the reflective surface(see Fig \ref{fig:setup}b). This collimated light is directed towards a set of two balanced photo detectors (BPD) (Thorlabs PDB210C) for interferometric detection depicted in Fig. \ref{fig:setup}c.  Using these BPDs, with a D-mirror in the path of one, the detection of motional degrees of freedom along $x'$, $y'$ and $z'$ can be maximized. \textcolor{black}{A phase locked loop parametric cooling scheme \cite{Jain:2016} is used to cool the particles mechanical degrees of freedom via optical means, while digitally implemented band-pass and derivative filters are used for electrostatic linear feedback as detailed in Section IV D.}

%#########################################################################################################################################
%#########################################################################################################################################

\section{\label{sec:results}Results}
\subsection{\label{sec:focusmoving}Characterization of Distance to Surface}
\begin{figure}
    \centering
    \begin{subfigure}{\linewidth}
        \centering
        \includegraphics[width=\linewidth]{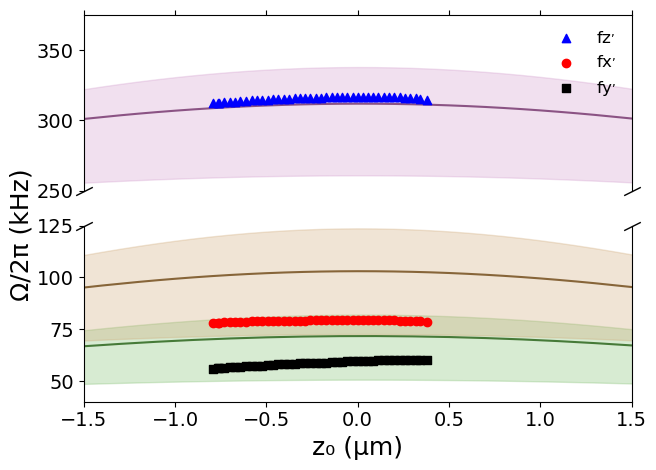}
        \label{fig:freqvspos2}
    \end{subfigure}
    
    \vspace{0.1cm} % Adjust spacing as needed

    \begin{subfigure}{\linewidth}
        \centering
        \includegraphics[width=\linewidth]{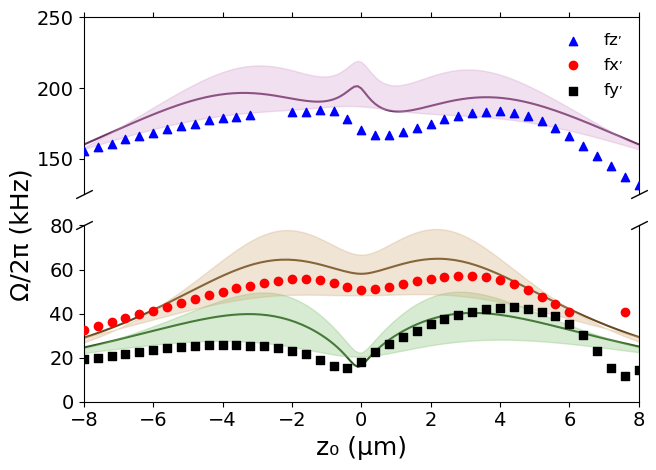}
        \label{fig:freqvspos1}
    \end{subfigure}
    
    \caption{Frequency versus focus location relative to the surface. The experimental data is shown as blue triangles for $f_{z'}$, red circles for $f_{x'}$ and black squares for $f_{y'}$. The shaded regions around each set of data represents the calculated values for the ``best" and ``worst" possible choices of laser power and beam waist. The solid line represents the expected values for these parameters. The top figure is for the case of the particle in the first potential well ($d_1 = 547$ nm), and the bottom for the case of the second potential well ($d_2 = 1.615\mu$m). This distinct quality of the two curves provides clear indication of the distance between the particle and the wall by simply measuring the COM motion as a function of focus location. }%For the upper bound of shaded region, the waist is determined from the fitting the beam waist to the Debye integrals for a tightly focused beam \textcolor{red}{cite principles of nanooptics} where $w_{x0} = 1.36 \, \mu m$, $w_{y0} = 1.46\mu$m and $P = 400$ mW.  For the lower bound,  $w_{0} = 1.8$ and $P=360$ mW. For the solid lines, the parameters values are $w_{0} = 1.51\mu$m and $P=380$ mW. The top figure is for the case of the particle in the first potential well ($d_1 = 547$ nm) where the data is taken with $50$ nm step size, and the bottom figure is for the case of a particle in the second potential well ($d_2 = 1.634\mu$m) with a step size of $400$ nm.}
    \label{fig:column_figures}
\end{figure}
By creating an optical lattice along the direction normal to the surface using the specified beam parameters, only two trapping sites exist. The distance of the lattice sites on the $z'$ axis are determined by the wavelength of the light used to create the lattice and the angle between the surface and the beam. In this case, light is incident at $45$ degrees relative to the line normal to the surface. At this angle, and the locations of the trap sites relative to the surface are approximately given by $(2n+1)\sqrt{2}\frac{\lambda}{4}$ where $n= \{0, 1\}$. In practice, there are small deviations from these values due to the Gouy phase and scattering force. %By modeling the optical potential, including the scattering force, with two orthogonal Gaussian TEM$_{00}$ mode beams with beam parameters specified in Fig 1a, the locations where a particle can be confined are located at $d_1 = 547$ nm and $d_2 =1.615$ nm away from the surface along $z'$.
The potential landscape experienced by a particle is very different depending on the trapping site, leading to clearly distinct trap frequencies for each potential well (see Fig. \ref{fig:column_figures}). 

To experimentally characterize the distance of the particle to the surface, %the distance along the $z$-direction is determined by measuring the COM as a function of the focus location relative to the trap minimum.  
a particle is trapped in one of the two sites and the focus is scanned over the area resulting in a change in trap frequency. Using the nano-positoning stage, the focus scans by the particle, reflects off the surface, and scans by the particle again resulting in a change in trap frequency at each focus location relative to the particle position. The shape of this curve will depend on which trapping site the particle is in as shown in Fig. \ref{fig:column_figures}. For a particle in the second trap site, 400 nm steps are taken (Fig. \ref{fig:column_figures} bottom). As expected from theory, there are local maxima which form at locations where the optical focus scans over the particle. For the particle in the first trap site, steps of $50$ nm are taken to have a finer resolution (Fig. \ref{fig:column_figures} top), and as expected from theory, there are no local maxima features. This is because the distance between the particle and the surface is much smaller than the waist ($d_1 << \omega_0$). \textcolor{black}{While we do not fully understand the reason for deviations from the theoretical model, it is possible that deviations occur due to effects such as spatially varying wavefront distortion and
changes in the reflected beam profile or polarization. Such perturbations can affect interference of the beam and its reflection that produces the optical potential. Further,} localized surface imperfections can generate intensity gradients which are especially pronounced close to the surface. These effects are not included in the simple model, and could \textcolor{black}{contribute to the} lack of a consistent trend around the expected frequency change for the different degrees of freedom.  \textcolor{black}{Additionally, optical aberrations or misalignment can cause asymmetry along the optical axis  and wavefront distortions which may contribute to a deviation from the theoretical model.} Without including such systematics into our model, most of the data still lie within the confidence bands. The solid colored line represents the expected value assuming a waist of $w_0 = 1.51$ $\mu$m found through fitting single beam spectra and typical measured laser power of $P=380$ mW. The shaded regions indicate an uncertainty band where the upper bound is calculated using waists $w_{x0} = 1.36$ $\mu$m and $w_{y0} = 1.46$ $\mu$m from fitting the Debye diffraction integrals and power $P=400$ mW. The lower bound takes parameters $w_{0} = 1.8$ $\mu$m and $P=360$ mW chosen based on typical objective misalignment and uncertainty in trapping power. Considering the deviations in the distance between the particle and the surface given by the upper and lower limits of the uncertainty band, and an uncertainty of $\pm 1$ degree in the angle of surface relative to the incident beam, bounds on the expected values are calculated. The distance between the particle and the surface is found to be $d_1 = 547^{+10}_{-10}$ nm for the first trap site and $d_2 = 1.615^{+0.019}_{-0.011}$ $\mu$m for the second trap site. This data experimentally validates the distinct trap frequencies in the two possible potential wells and the calculated distance between the particle and the surface. Additionally, the ability to move the reflective surface along $x$, $y$ and $z$ directions with such resolution and accuracy provides a method for tuning the frequencies of the COM motion and doing force scanning experiments.

%#########################################################################################################################################
%#########################################################################################################################################

\subsection{\label{sec:polarization}Tuning Frequencies via Laser Polarization}
\begin{figure}[h!]
   \centering
   \includegraphics[width=\linewidth]{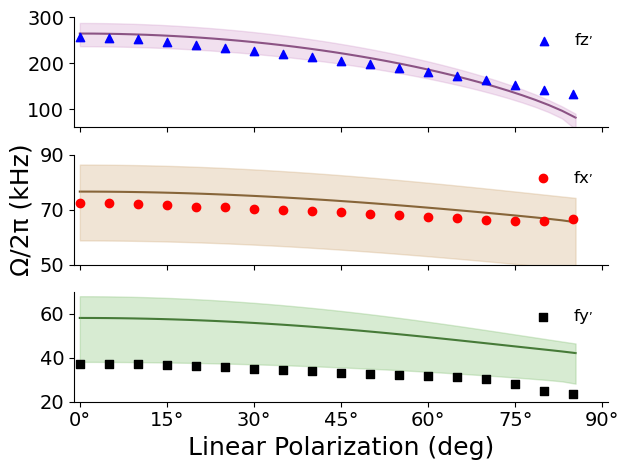}
   \caption{Frequency versus linear polarization of the trapping light for a particle trapped in the second potential well ($d_2 = 1.634\mu$m). The experimental data is shown as blue triangles for $f_{z'}$, red circles for $f_{x'}$ and black squares for $f_{y'}$. The shaded regions around each set of data represents the calculated values for the ``best" and ``worst" possible choices of laser power and beam waist. The solid line represents the expected experimental for these parameters. It is clear that by turning the linear polarization closer to $90$ degrees, the interference between the incident and reflected beam decreases causing a decrease in potential depth. Note that the qualitative shape of these depends slightly on the location of the focus.} %For the upper bound of the shaded region, waist values $w_{x0} = 1.36 \, \mu m$, $w_{y0} = 1.46\mu$m and laser power $P = 750$mW are chosen. For the lower bound, $w_0 = 1.8$ $\mu$m and $P = 625$ mW. For the solid line, values $w_0 = 1.51$ $\mu$m and $P = 700$ mW are chosen.}
   \label{fig:freqvspol}
\end{figure}
To form an optical standing wave potential, two beams must interfere. For a standing wave trap formed by a tilted mirror, the polarization of the light reflected from the tilted surface will not always be equal to the polarization of the incident light. Thus the amount of interference generating the optical potential depends on the laser polarization. This is not the case in retro-reflected optical standing wave traps \cite{Montoya:22}.  Furthermore, the ability to change the potential depth, and thus the trap frequencies, by simply adjusting the polarization of the trapping light \textit{in-situ} demonstrates the versatility of a tilted standing wave trap. To characterize this effect, the trap frequencies of a particle in the second potential well are measured as a function of linear polarization where zero degrees corresponds to polarization along the $x$ axis (see Fig. \ref{fig:freqvspol}). For this study, a higher laser power of approximately $700$mW is used so that the particle would stay trapped for linear polarizations closer to $90$ degrees (aligned along the y - axis). Additionally, this data is collected for the case where the optical focus is in the vicinity of the particle in the second potential well (rather than on the surface). As expected, Fig. \ref{fig:freqvspol} shows that for $0$ degrees polarization, the trap frequencies are highest indicating maximal interference. As the linear polarization is increased, the frequencies decrease due to less interference. For the expected curve and its uncertainty band, the same waists as Fig. \ref{fig:column_figures} are chosen, but with laser powers of $750$ mW for the upper bound, $625$ mW for the lower bound and $700$ mW for the expected curve. In the case of the $f_{z'}$ data, the frequency is changed by over $100$ kHz simply by changing the linear polarization on the trapping beam via a waveplate. This capability to dramatically tune the trap frequencies of a mechanical oscillator provides unique opportunities for resonant force detection over a range of frequencies.

\subsection{\label{sec:simulations} Langevin Simulation of Particle Dynamics}
The particle dynamics are governed by the Langevin equations
\begin{equation}
d\vec{v} = \left[ \frac{1}{4M} \alpha' \nabla |\vec{E}|^2 - \frac{\mu_0 \omega}{M} \alpha'' \langle \vec{S} \rangle - \Gamma_{\text{tot}} \vec{v} \right] dt + A\, dW
\label{eq:eom1}
\end{equation}
\begin{equation}
d\vec{r} = \vec{v} dt
\label{eq:eom2}
\end{equation}
\noindent where $M$ is the particle mass, $\alpha = \alpha' + i\alpha''$ is the complex polarizability, $\vec{E}$ is the total electric field, $\langle \vec{S} \rangle$ is the time-averaged Poynting vector, and $\Gamma_{\text{tot}}$ is the total damping. In the pressure regime we study, the damping and photon recoil heating due to the trap laser can be neglected. The stochastic force is represented by the Wiener increment $dW$ with amplitude $A = \sqrt{2 k_B T_{\text{CM}} \Gamma_{\text{tot}} / M}$, where $T_{\text{CM}}$ is the center-of-mass temperature. 

To better understand our system and compare experimental results with theoretical predictions, we performed Langevin simulations using Northwestern’s QUEST cluster. The equations of motion \ref{eq:eom1},\ref{eq:eom2} were numerically integrated for a 45 degree tilt and a reflected Gaussian TEM$_{00}$ mode with the following parameters: beam waists $w_{x0} = 1.365\,\mu$m, $w_{y0} = 1.464\,\mu$m, Rayleigh length $z_{R} = 3.683\,\mu$m, power $P_{0} = 0.4\,W$, silica sphere radius $R = 85\,nm$, air pressure $p = 2$ Torr, and gas temperature $T = 300$ K.
\subsubsection{Adiabatic Transition and Trapping in the Second Potential Well}

\begin{figure}
    \centering
    \includegraphics[width=1\linewidth]{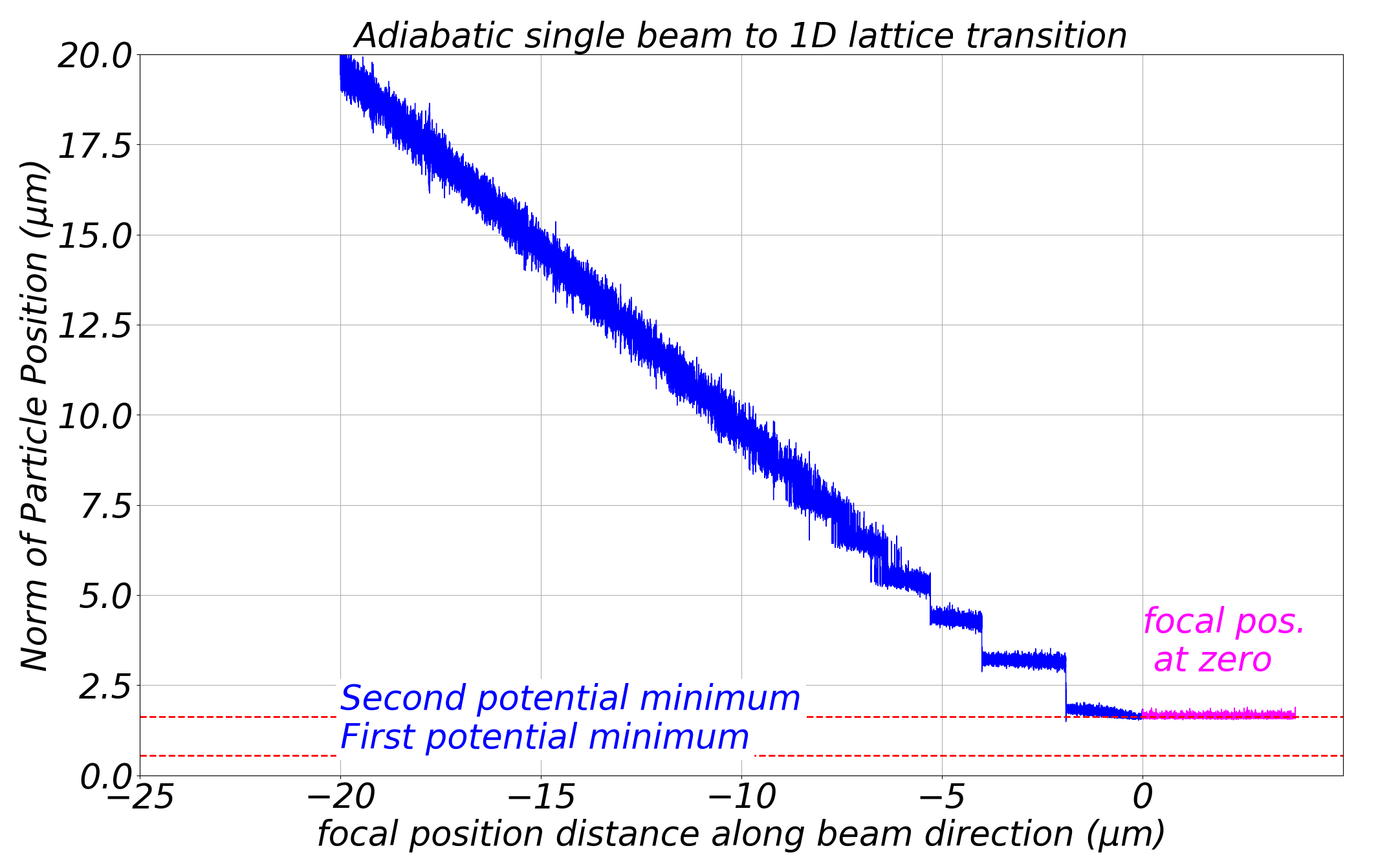}
    \caption{Simulated particle dynamics during an adiabatic transition from a single-beam trap to a one-dimensional optical lattice. The focal position of the laser beam is continuously shifted along the propagation direction at a velocity of $38.3 \, \mu$m/s. As new intermediate lattice sites form (compare Fig. \ref{fig:column_figures}), the particle sequentially jumps into them, ultimately settling in the second potential well—the farthest stable site from the mirror in this configuration. A relaxation period of $100 \, \mu$s is included at the end, with the focal position held at the mirror surface.}
    \label{fig:adiabatic}
\end{figure}

In our first simulation, we confirm the experimental observation that the particle consistently gets trapped in the second potential minimum upon transition from the single beam tweezer. Figure \ref{fig:adiabatic} illustrates the simulated particle dynamics during an adiabatic transition from a single-beam trap to a one-dimensional optical lattice. The laser beam focus is continuously shifted along the propagation direction (compare Fig. \ref{fig:potentialcontour1}) at a velocity of $38.3\,\mu$m/s. As new intermediate lattice sites form (see Fig. \ref{fig:potentialcontour1}), the particle sequentially jumps into them, ultimately stabilizing in the second potential well—the farthest stable site from the mirror in this setup.

A relaxation period of $100\,\mu$s is included at the end, with the focal position held at the mirror surface. Notably, the intermediate lattice sites, when the focus is not at the reflecting surface, are not necessarily spaced by $\sqrt{2}\lambda/2$. This deviation arises due to phase shifts from Gouy phase and wavefront curvature mismatches between the incident and reflected beams.

\begin{figure}
    \centering
    \includegraphics[width=1\linewidth]{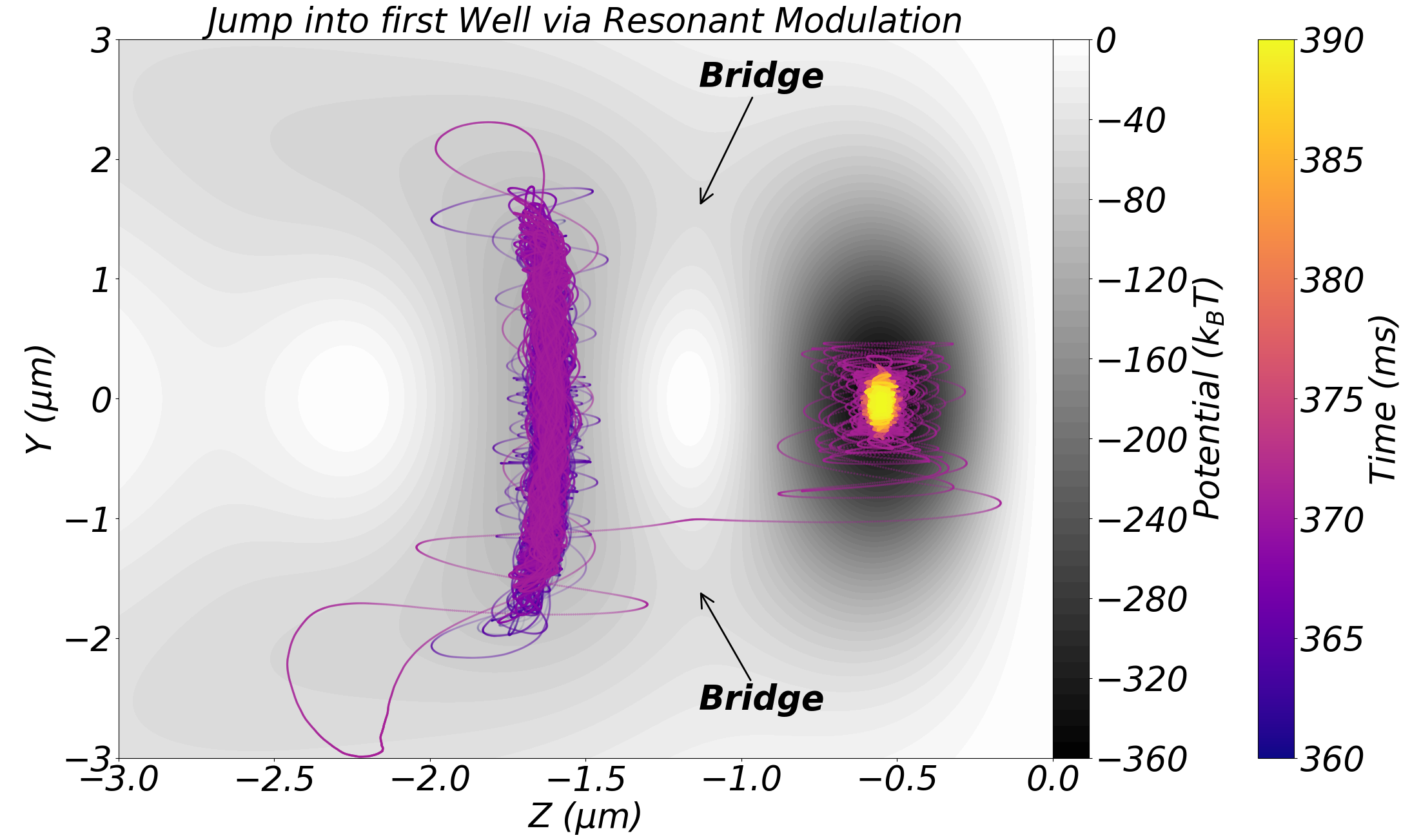}
    \caption{Particle trajectory for a controlled transition from the second potential well to the first via resonant modulation of the laser power. The laser intensity is modulated sinusoidally at the resonance frequency $f_{y'}=21,137.51$ Hz of the second well, inducing a periodic variation in the scattering force along the $y'$-direction. This leads to resonant heating of the particle’s motion, gradually redistributing energy across all degrees of freedom. As the modulation depth $\beta$ increases linearly as $\beta = 2 t$ (for $t$ between 0 and 0.5 sec), the energy eventually reaches a threshold at $t=371$ ms ($\beta = 0.74$), triggering the transition. The particle crosses one of the two lower-energy "bridges" ($\sim 44 k_B T$ below the escape barrier), making the transition highly probable. The trajectory is shown only around the transition time to maintain clarity and avoid oversaturation.}
    \label{fig:jump}
\end{figure}

\subsubsection{Controlled Transition via Resonant Modulation}

\begin{figure*}
    \centering
    \includegraphics[width=\textwidth]{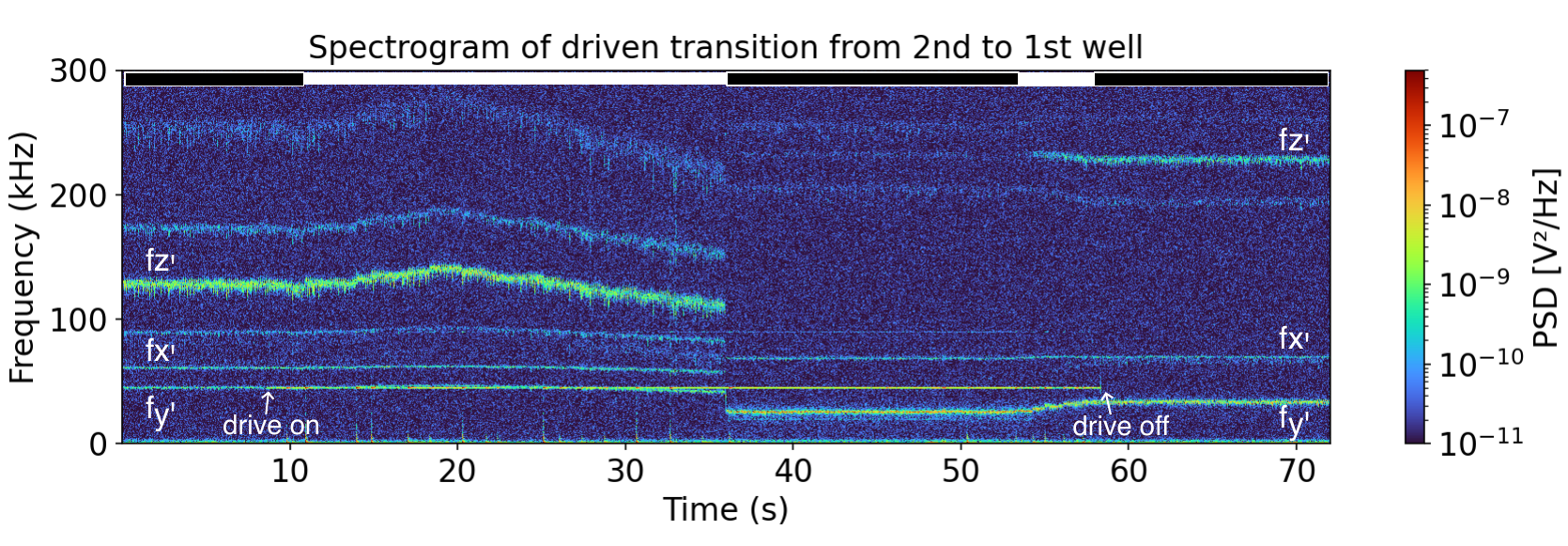}
    \caption{\textcolor{black}{Spectrogram of experimental data demonstrating the driven transition from the second to first potential well.  The tilted mirror is translated toward the particle using an in vacuum stage. At time zero, the particle is in the second potential well. The black bars on top indicate times when the position of the focus is stationary, while the white bars indicate times when the focus is moving closer to the surface in discrete steps. Around $20$ s there is a maximum in the frequencies (see also Fig. \ref{fig:column_figures}) indicating that the focus is on top of the particle in the second well, and past this time the focus extends between the second well and the surface. A driving frequency from a function generator into the EOM supplies a resonant modulation to the $f_{y'}$ peak at $45$ kHz with a modulation depth of $0.22$. Around $36$ s, with the drive on, an abrupt transition occurs and the frequencies become consistent with expected values for the first potential well. Around $55$ s, the focus is moved closer to the surface briefly to adjust the frequencies and improve signal to noise.}}
    \label{fig:driventransition}
\end{figure*}

Although the particle naturally settles in the most distant stable potential well, this is not a practical limitation.  By applying parametric modulation (heating) at a resonance frequency, the particle can be transferred to the first potential well. In this method, a sinusoidal modulation of the laser power (e.g., using an electro-optic modulator, EOM) drives the scattering force, effectively heating the particle’s motion. Linear modulation is preferable due to its reduced noise. Figure \ref{fig:jump} depicts the particle trajectory during this controlled transition.

The laser intensity is modulated sinusoidally at the resonance frequency $f_{y'} \approx 21.1$ kHz of the second well, inducing periodic variations in the scattering force along the $y'$-axis. This process leads to resonant heating, gradually redistributing energy across all degrees of freedom. As the modulation depth $\beta$ increases linearly as $\beta = 2t$ (for $t$ between 0 and 0.5 sec), the energy eventually surpasses a critical threshold at $t = 371$ ms ($\beta = 0.74$), triggering the transition. The particle then crosses one of the two lower-energy ``bridges" (approximately $44 k_B T$ below the escape barrier), making the transition highly probable. To maintain clarity, only the trajectory near the transition time is shown, avoiding oversaturation.

\textcolor{black}{To experimentally test the ability to drive the particle from the second to the first well, Fig \ref{fig:driventransition} shows the trap frequencies as a function of time while modulating the potential at the same frequency as $f_{y'}$. The laser polarization was chosen to be elliptical to reduce the barrier between the wells and ease the transition into the first well. The laser power was chosen to be approximately $700$ mW to maintain sufficient trap depth in the second well. The black and white bars on the top of the spectrogram represent time intervals where the focus is stationary (black) and when it is being moved towards the surface in discrete steps (white). Initially, the particle is in the second well, and as the focus traverses the second potential well there is a maximum in frequencies seen particularly clearly on the $f_{z'}$ peak around $20$ s (see also Fig. \ref{fig:column_figures}). After this, the focus is between the second well and the surface. A function generator is used to resonantly drive the $f_{y'}$ peak at $45$ kHz with a modulation depth of $0.22$. In the figure, the start and stop of the drive is indicated by ``drive on" and ``drive off", and an abrupt transition in the trap frequencies is seen around $36$ s. These frequencies after $36$ s are consistent with the particle being in the first potential well. The focus is then briefly moved towards the surface again to adjust frequency and signal to noise. %This spectrogram illustrates the fact that the particle in this optical potential does not follow the position of the focus as it would in systems using surfaces that are mostly optically transparent \cite{Novotnymembrane, Winstone2018}, and furthermore demonstrates the ability to controllably drive the particle from the second to the first well in a manner consistent with the simulation.
}

\begin{figure}[h!]
   \centering
   \includegraphics[width=\linewidth]{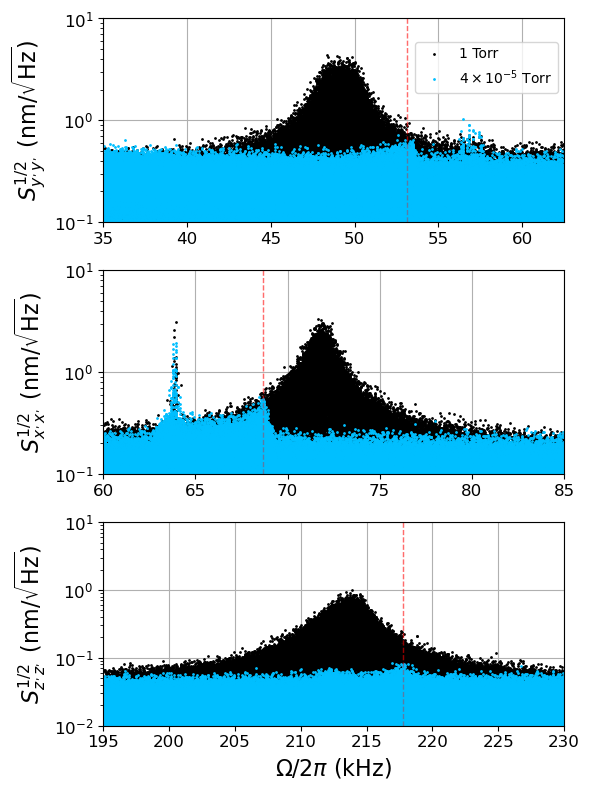}
   \caption{\textcolor{black}{PSD for a 170nm diameter silica sphere in the first potential well for chamber pressure of $1$ Torr (black) and $4.0\times10^{-5}$ Torr (blue).  All three axes are being cooled simultaneously with a single tungsten probe tip as an electrode. The sharp peak around $64$ kHz is detector noise and is present even with the particle absent. The distorted line shape in the $x'$ cooling data is suspected to be due to this detector noise being fed back onto the particle motion. The vertical dashed red lines indicate the approximate location of the frequency peaks after being cooled.}}
   \label{fig:3Dcoolingelectrode}
\end{figure}

\subsection{\label{sec:feedback} 3D Feedback Cooling}
\begin{figure}[h!]
   \centering
   \includegraphics[width=\linewidth]{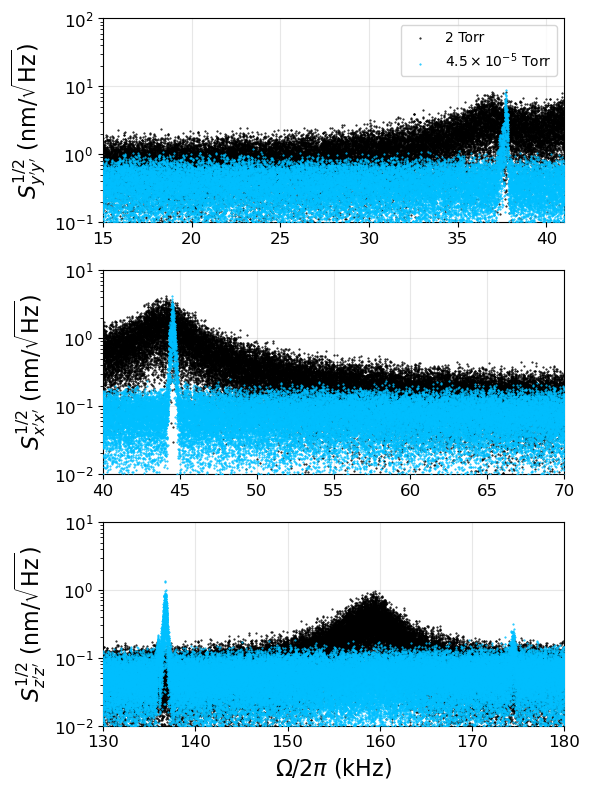}
   \caption{PSD for a 170nm diameter silica sphere in the second potential for chamber pressure of 2 Torr (black) and $4.5 × 10^{-5}$ Torr (blue). The y and z axes are being cooled simultaneously, while the x axis is from a subsequent data set with different cooling efficiency for y and z.}
   \label{fig:3Dcoolingparametric}
\end{figure}

For many levitated optomechanical experiments, including ultrasensitive force measurements, it is necessary to cool the COM motion of the nanoparticle so that it remains trapped in high vacuum conditions. \textcolor{black}{Linear feedback with an electrode has become a commonly used method for cooling the COM motion of charged levitated nanoparticles, and has even been used for ground state cooling \cite{groundstate, groundstatecryo}. To demonstrate compatibility of our apparatus with this method, Fig. \ref{fig:3Dcoolingelectrode} shows data of all three COM degrees of freedom being cooled via a single tungsten probe tip as an electrode place approximately $5$mm away from the particle. This data is taken for a particle in the first potential well ($d_1 = 0.547\mu$m) from the conducting surface, using the same laser power and polarization as in Fig \ref{fig:driventransition}. By modeling the electric field of the electrode in the vicinity of the grounded conducting surface in COMSOL and driving the particle motion slightly below resonance via a function generator, we estimate the magnitude of the net surface charge to be $2.02\pm 0.15$. \textcolor{black}{See Appendix \ref{sec:appendixC} for more details.} The feedback signal for each degree of freedom is generated by applying a second order band-pass filter followed by a derivative filter.  These three feedback signals are then summed and amplified by a high voltage amplifier. The estimated temperatures associated with the COM mechanical degrees of freedom at $4 \times 10^{-5}$ Torr shown in Fig. \ref{fig:3Dcoolingelectrode} are $T_{y'} \approx 2.83$ K, $T_{x'} \approx 3.86$ K and $T_{z'} \approx 0.97$ K.} We attribute the distorted line shape in for the $x'$ motion to the detector noise peak around $64$ kHz which is being fed back onto the particle motion while feedback cooling.

This apparatus is also compatible with optical parametric feedback cooling methods where a modulation of the trapping potential is used to reduce the particles mechanical motion. \textcolor{black}{This method does not rely on the particle having a non-zero surface charge.} Results of parametric feedback cooling of all three COM motions is demonstrated here for a particle located in the second potential well from the surface ($d_2 = 1.615\mu$m). This is done by using a phased locked loop (PLL) to generated a feedback signal at two times the trap frequency for each degree of freedom. Using an EOM as an amplitude modulator, the feedback signal for each motion is used to parametrically cool the particle motion\cite{parametric}\cite{Jain:2016}. Using this method, sufficient 3D parametric feedback cooling was achieved in order to reliably keep particles trapped at $1.3 \times 10^{-5}$ Torr.  The pressure of $1.3 \times 10^{-5}$ Torr is the limit of the turbo pump currently being used in the system, and in principle higher vacuum should be possible with the same apparatus. The data in Fig. \ref{fig:3Dcoolingparametric} shows the data collected for a particle which is initially in a 2 Torr environment and eventually is held at $4.5 \times 10^{-5}$ Torr with small amounts of feedback cooling applied. The estimated temperatures associated with the COM motional degrees of freedom in high vacuum shown in Fig. \ref{fig:3Dcoolingparametric}. are $T_{y'} \approx 61$ K, $T_{x'} \approx 28$ K and $T_{z'} \approx 30$ K, with associated cooling rates $\Gamma_{y'} \approx 156$ Hz, $\Gamma_{x'} \approx 239$ Hz and $\Gamma_{z'} \approx 364$ Hz. \textcolor{black}{Although we expect that the cooling performance could in principle be improved by increasing the signal-to-noise ratio of the displacement detection and optimizing the parameters of the parametric feedback circuit and phase locked loop configuration as has been done in other parametric cooling experiments far away from surfaces by other groups e.g. in Ref. \cite{Jain:2016},} demonstrating the capability to feedback cool all three degrees of freedom of a levitated nanoparticle at high vacuum less than a micron away from a conducting surface shows promise for future goals of doing ultrasensitive force measurements in a cryogenic ultrahigh vacuum environment. %Demonstrating the capability to feedback cool all three degrees of freedom of a levitated nanoparticle at high vacuum less than a micron away from a conducting surface shows promise for future goals of doing ultrasensitive force measurements in a cryogenic ultrahigh vacuum environment.   

\section{\label{sec:discussion} Discussion and Outlook}

In this work, we have proposed and experimentally demonstrated a novel optical trapping configuration for dielectric particles near a reflecting surface. An analytical model, based on Gaussian beam optics, has been verified through experimental data. While we observe good qualitative agreement, slight deviations arise due to uncertainty in the precise beam waist, wavefront distortions, and imperfections in the gold-coated mirrors. Additionally, we have validated our understanding of the particle dynamics by numerically solving the Langevin equations of motion.

Our approach offers several distinct advantages. Firstly, when tightly focused beams cross at an angle, only a finite number of potential wells form in the overlapping region, reducing complexity and enabling controlled trapping.
    %\item The transition from a single-beam trap to a one-dimensional optical lattice occurs continuously, as the light is not retroreflected.
    %\item Particles are loaded reliably without uncontrolled transient behavior or significant losses.
In addition, the trapping frequency and position of the potential wells can be tuned by adjusting the polarization, incidence angle and relative position between focus and particle. Thirdly, particles can be deterministically placed into the nearest potential well via linear or parametric heating or by increasing the incidence angle to allow only a single stable minimum.

Finally, we have demonstrated the ability to stably trap \textcolor{black}{and cool $170$ nm silica spheres at (sub-)micron range from a tilted metallic surface using two distinct methods. It has been shown that the particles do not need to be neutral in order to be stably trapped near the surface, and it is additionally possible to utilize the charge for feedback cooling.} In future work, we aim to extend trapping capabilities to fully cryogenic environments and explore direct trapping at thin membranes, enabling novel surface interaction studies. We also note that our apparatus is compatible with \textcolor{black}{optical} linear feedback cooling \cite{GeraciZepto}. Our results open new opportunities for precision force measurements at sub-micron distances \textcolor{black}{with neutral and charged particles}. This platform could be useful for tests of non-Newtonian gravity, short-range fifth forces, and experimental studies of out-of-equilibrium Casimir-Polder forces. Additionally, by trapping a charged particle near the tilted surface, this could enable surface potentiometry. Furthermore, if equipped with suitable scanning capabilities, it offers a robust setup for ultra-sensitive opto-levitated surface microscopy.

%These advancements will further enhance the potential of optically levitated systems for fundamental physics and precision sensing applications.

%#########################################################################################################################################
%#########################################################################################################################################

\section{Acknowledgements}
A. Grinin and A. Geraci are thankful to the Humboldt Society for the Feodor-Lynen postdoctoral fellowship granted to A. Grinin.  A. Geraci acknowledges support from NSF grants PHY-2409472 and PHY-2111544, DARPA, the John Templeton Foundation, the W.M. Keck Foundation, the Gordon and Betty Moore Foundation Grant GBMF12328, DOI 10.37807/GBMF12328, the Alfred P. Sloan Foundation under Grant No.\ G-2023-21130
%\subsubsection{Wide text (A level-3 head)}

\appendix
\section{Optical Potential for arbitrary Angle of Incidence} \label{sec:appendixA}

% Content for Appendix A

In this appendix we present a more generalized formula for the optical potential. It includes an arbitrary incidence angle and arbitrary phase due to reflection (e.g.\ metallic mirrors). Further, we show how the more accurate Debye integral approach differs from its Gaussian‐beam approximation and justify our choice to model the optical potential with Gaussian beams.
For a fully general incidence angle and mirror phase, the optical potential can be written immediately as

\onecolumngrid
\begin{equation}
U(\vec{r})
= -\frac{\alpha'}{4}\,
\Bigl[
  \vec{\mathcal{E}}_{\rm inc}^2(\vec{r})
  + \vec{\mathcal{E}}_{\rm ref}^2(\vec{r})
  + 2\,\vec{\mathcal{E}}_{\rm inc}(\vec{r})\cdot \vec{\mathcal{E}}_{\rm ref}(\vec{r})\,
    \cos\!\bigl(\phi_{\rm ref}(\vec{r})-\phi_{\rm inc}(\vec{r})\bigr)
\Bigr].
\end{equation}
\twocolumngrid

\noindent where the Gaussian envelopes are
\begin{equation}
\vec{\mathcal{E}}_{\rm inc}(\vec{r})
= E_0\,\frac{w_{0u}\,w_{0v}}{w_u(s_{\rm inc})\,w_v(s_{\rm inc})}
  \exp\!\Bigl[-\tfrac{\rho_{u,\rm inc}^2}{w_u^2(s_{\rm inc})}
               -\tfrac{\rho_{v,\rm inc}^2}{w_v^2(s_{\rm inc})}\Bigr],    
\end{equation}
\begin{equation}
\vec{\mathcal{E}}_{\rm ref}(\vec{r})
= E_0\,\frac{w_{0u}\,w_{0v}}{w_u(s_{\rm ref})\,w_v(s_{\rm ref})}
  \exp\!\Bigl[-\tfrac{\rho_{u,\rm ref}^2}{w_u^2(s_{\rm ref})}
               -\tfrac{\rho_{v,\rm ref}^2}{w_v^2(s_{\rm ref})}\Bigr].    
\end{equation}

The corresponding phases are
\begin{equation}
\phi_{\rm inc}(\vec{r})
= k\,s_{\rm inc}
  + \frac{k\bigl(\rho_{u,\rm inc}^2+\rho_{v,\rm inc}^2\bigr)}{2\,R(s_{\rm inc})}
  - \xi\bigl(s_{\rm inc}\bigr),    
\end{equation}
\begin{equation}
\phi_{\rm ref}(\vec{r})
= k\,s_{\rm ref}
  + \frac{k\bigl(\rho_{u,\rm ref}^2+\rho_{v,\rm ref}^2\bigr)}{2\,R(s_{\rm ref})}
  - \xi\bigl(s_{\rm ref}\bigr)
  + \phi_{\rm mir}.    
\end{equation}

Here we have defined the beam‐center coordinates along each axis as
\begin{equation}
s_{\rm inc}
= \hat{\textbf{k}}_{\rm inc}\!\cdot\!\bigl(\vec{r}-\vec{r}_{0,\rm inc}\bigr),
\quad
\vec{\rho}_{\rm inc}
= \bigl(\vec{r}-\vec{r}_{0,\rm inc}\bigr) - s_{\rm inc}\,\hat{\textbf{k}}_{\rm inc},    
\end{equation}
\begin{equation}
s_{\rm ref}
= \hat{\textbf{k}}_{\rm ref} \cdot\!\bigl(\vec{r}-\vec{r}_{0,\rm ref}\bigr),
\quad
\vec{\rho}_{\rm ref}
= \bigl(\vec{r}-\vec{r}_{0,\rm ref}\bigr) - s_{\rm ref}\,\hat{\textbf{k}}_{\rm ref}.    
\end{equation}
Where $\hat{\textbf{k}}_{\rm inc} = \vec{k}_{inc}/k$ is the unity vector in the direction of the k-vector of the incident beam and the $\hat{\textbf{k}}_{\rm ref}$ correspondingly. The waist positions \(\vec{r}_{0,\rm inc}\) and \(\vec{r}_{0,\rm ref}\) are related as mirror images of each other. Finally, the beam parameters along each propagation direction are $w_i(s)=w_{0i}\sqrt{1+\bigl(s/z_R\bigr)^2}$, $R(s)=s\Bigl[1+\bigl(z_R/s\bigr)^2\Bigr]$, $\xi(s)=\arctan\bigl(s/z_R\bigr)$
\newline
In the special case of 45° incidence on a mirror with the waist positioned at the reflecting surface, these reduce to
$s_{\rm inc}=z,\quad \rho_{u,\rm inc}=x,\quad \rho_{v,\rm inc}=y$, $s_{\rm ref}=y,\quad \rho_{u,\rm ref}=x,\quad \rho_{v,\rm ref}=z$ 
recovering exactly the expressions in Sec.~\ref{sec:theory}.

\section{Comparison of Debye Diffraction Integral and Generalized Gaussian Beam Model} \label{sec:appendixB}
\begin{figure}[]
    \centering
    \includegraphics[width=1\linewidth]{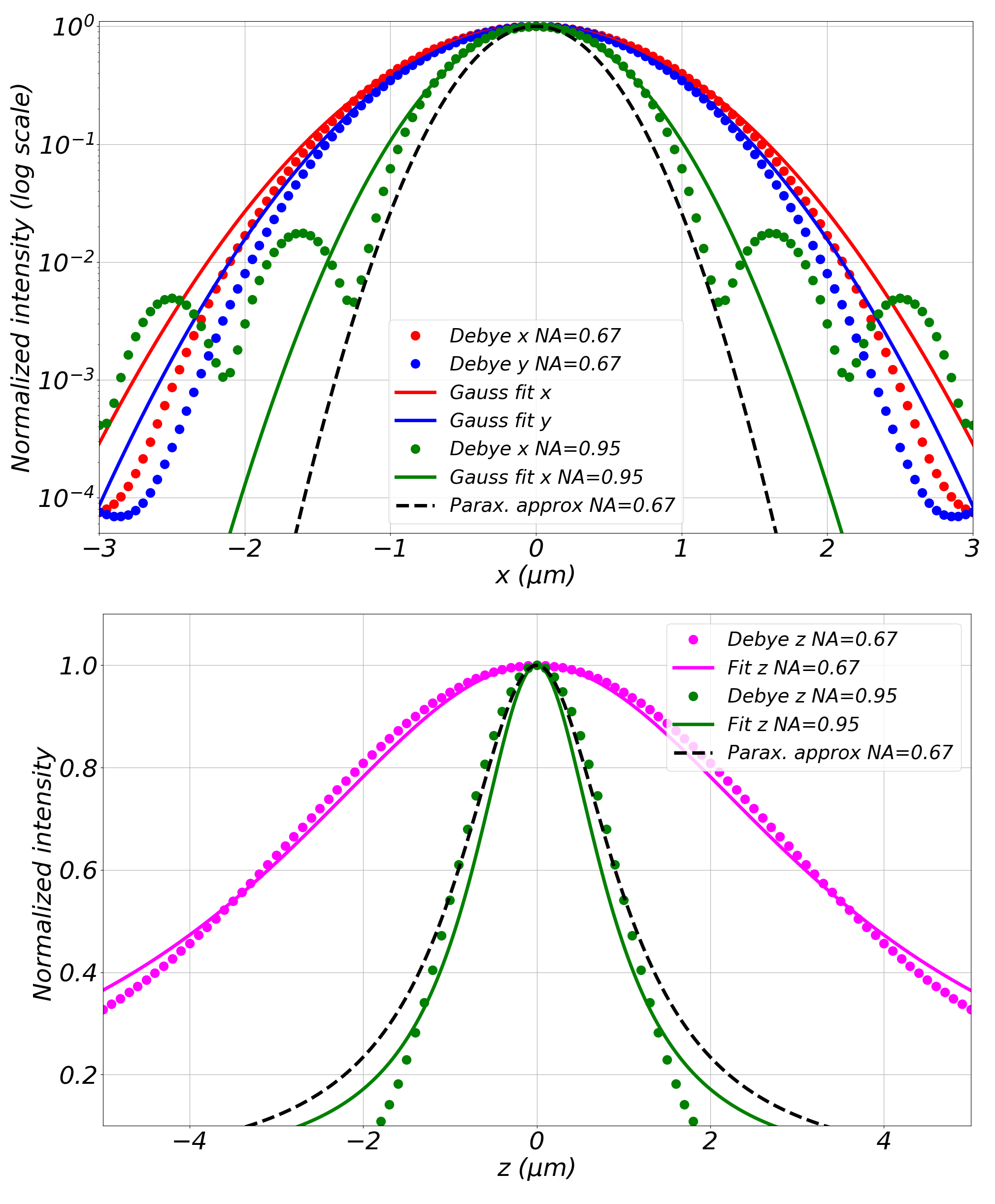}
    \caption{Normalized intensity as a function of the radial position x or y. Debye integral solutions together with a Gaussian fit (blue x-axis, red y-axis dots -data, solid line - fit, NA = 0.67, filling factor $f_0=0.56$). The lower plot shows the z-axis Debye integral and fit. Even for NA = 0.95, filling factor $f_0=2$ (green dots) the deviations from a Gaussian fit (solid green) around the peak are moderate. The dashed line shows the naive application of the paraxial approximation for an NA=0.67 objective.}
    \label{fig:debye_gauss}
\end{figure}

\textcolor{black}{To estimate the waist of our objective lens (OptoSigma PAL-50-NIR-HR-LC00), we used both a best-case theoretical estimate as well as measurement data. The theoretical best-case estimate (assuming no aberrations i.e. aplanatic lens) is obtained by calculating the intensity profile at the focal plane using the Debye integral approach. For an NA=0.67 one can fit the diffraction limited intensity profile with high accuracy with a Gaussian TEM$_{00}$ mode, where instead of using a paraxial waist and Rayleigh length, the fitted parameters are used. The focal length and the filling factor are 4mm and 0.6 correspondingly. This corresponds to the best case, i.e. aberration free and with perfect alignment into the objective. \newline
We also estimated the waist experimentally using a scanning slit beam profiler (Dataray Beam'R2) and found that while a diffraction limited waist can be obtained for best alignment $w_x\approx 1.3\mu$m and $w_y\approx 1.4\mu$m in accordance with the Debye Integral estimate, a typical alignment produces a waist of  $1.5\mu$m (expected value) and can be as large as $1.8\mu$m (worst-case without realignment for our experimental conditions). We verified the expected value by fitting the observed single-beam trap frequencies (i.e. with the particle far from the tilted mirror) given the measured beam power.  
}
\newline
Figure \ref{fig:debye_gauss} shows the numerical calculation of the Debye diffraction integral (dots) for the objective used in the experiment (NA=0.67, filling factor $f_0=0.56$, effective focal length f=4 mm) together with its Gaussian fit for the x-axis (polarization direction, red solid line, upper plot), the y-axis (blue, upper plot) and the z-axis (magenta, lower plot). We also evaluate and compare to the Gaussian fit approximation a very tightly focused beam of NA=0.95, filling factor $f_0=2$. For the z-axis we used the longitudinal intensity envelope fitting function $1/(1+(z/z_R)^2)$ with $z_R$ as independent parameter. As can be seen from the figure, the generalized (waists and Rayleigh length separate parameters) Gaussian beam model fits the Debye integral very accurately for all but very high numerical apertures and far away from the focal point. 

\section{Nanoparticle Charge Calibration} \label{sec:appendixC}
\textcolor{black}{To determine the charge of the levitated nanoparticle which is electrostatically feedback cooled in Fig. \ref{fig:3Dcoolingelectrode}, we first get a voltage-to-distance calibration of the $z'$ motion of the particle. This is done by fitting a Lorentzian function to the $f_{z'}$ peak in the power spectrum at a pressure of 1 Torr. At this pressure, we can assume the particle is in thermal equillibrium with its environment ($T = 293$ K) \cite{Montoya:22, GeraciAtto}. The voltage-to-displacement calibration factor is $\alpha = \sqrt{k_B T/m\omega_0^2V^2 }=6.22 \times 10^{-6}$ m/V. A $10$ V is applied to the electrode is at $200$ kHz driving the particles motion via the Coulomb force. %The $z'$ displacement of the particle due to the Coulomb force from the electrode is found by finding the average amplitude of the displacement at the driving frequency. 
In a $0.1$ ms portion of the time series at the driving frequency $f_{d}$, the average displacement is $\Delta z'(f_{d}) = \alpha \times 1.29$ $\mu$V $=8.02\times 10^{-11}$ m. A COMSOL model is used to estimate the magnitude of the electric field along the $z'$ direction at the particles location relative to the surface. We find this value to be $E_{z'} = 231$ V/m.  With the fitted value for the frequency $f_{z'} = 214$ kHz and the calculated particle mass $M = 5.14 \times 10^{-18}$ kg, we estimate the charge by equating $k\Delta z'(f_{d}) = neE_{z'}$ where $k =9.26 \times 10^{-6}$ N/m is the optical trap stiffness along $z'$, $e = 1.6 \times 10^{-19}$ C is the charge of the electron and $n$ is the number of charges. We determine a value of $n = 2.02 \pm 0.15$ where the uncertainty comes primarily from the variance in the average value of $\Delta z'(f_{d})$.}

\begin{comment}
\begin{table}[tbp]
    \centering
    \caption{Taking the calibration factor $\alpha$, the displacment of the particle $\Delta z'$ due to the force exerted by the electrode, and the value of the electric field magnitude along $z'$ modeled in COMSOL an estimate of the net charge on the surface of the particle is calculated to be $2.02 \pm 0.15$.}
    \label{tab:chargecal}
    \begin{tabular}{ll}
    \toprule
    Parameter & Value \\
    \midrule
    $\alpha$   & $6.22 \times 10^{-6}$ m/V   \\
    $\Delta z'$    & $8.02\times 10^{-11}$ m   \\
    $E_{z'}$    & $231$ V/m   \\
    $k$    & $9.26 \times 10^{-6}$ N/m   \\
    $n$  & $2.02 \pm 0.15$  \\
    \bottomrule
    \end{tabular}
\end{table}
\end{comment}

%\section{Simulation Details} \label{sec:appendixB}
% Content for Appendix B
%\subsection{\label{sec:citeref}Citations and References}
%A citation in text uses the command \verb+\cite{#1}+ or
%\verb+\onlinecite{#1}+ and refers to an entry in the bibliography. 
%An entry in the bibliography is a reference to another document.

%\subsubsection{Citations}
%\section*{}
%\noindent *These authors contributed equally to this work.
\bibliography{apssamp_main2}% Produces the bibliography via BibTeX.

\end{document}